\documentclass[aps,showpacs]{revtex4}
\usepackage{amssymb}
\usepackage{graphicx}
\usepackage[english]{babel}
\usepackage{indentfirst}
\usepackage{amsxtra}
\usepackage{amsmath}
\usepackage{supertabular}
\usepackage{multirow}
\usepackage [mathcal]{eucal}
\def\beq{\begin{equation}}
\def\eeq{\end{equation}}
\def\beqn{ \begin{eqnarray} }
\def\eeqn{ \end{eqnarray} }

\def\s1s2{{ \boldsymbol{\sigma}(1) \cdot \boldsymbol{\sigma}(2) }}
\def\t1t2{{ \boldsymbol{\tau}(1) \cdot \boldsymbol{\tau}(2)  }}
\begin{document}
\noindent
\title{
Self-consistent continuum random phase approximation calculations of 
        $^4$He electromagnetic responses
}
\author{V. De Donno$^{\,1,2}$, M. Anguiano$^{\,3}$, G. Co'$\,^{\,1,2}$,  
and  A. M. Lallena$^{\,3}$}
\affiliation{
 \mbox {1) Dipartimento di Fisica, Universit\`a del Salento,
Via Arnesano, I-73100 Lecce, ITALY} \\
\mbox {2) INFN, Sezione di Lecce, Via Arnesano, I-73100 Lecce, ITALY } \\
\mbox {3) Departamento de F\'\i sica At\'omica, Molecular y
  Nuclear,} \\ 
\mbox {Universidad de Granada, E-18071 Granada, SPAIN } \\
}
\date{\today}

\bigskip

\begin{abstract} 
We study the electromagnetic responses of $^4$He within the framework
of the self-consistent continuum random phase approximation theory.
In this approach the ground state properties are described by a
Hartree-Fock calculation.  The single particle basis constructed in
this manner is used in the calculations of the continuum responses of
the system. Finite-range interactions are considered in the calculations.
We compare our results with photon absorption cross sections and
electron scattering quasi-elastic data. From this comparison, and also
from the comparison with the results of microscopic calculations, we
deduce that our approach describes well the continuum excitation.
\end{abstract}

\bigskip
\bigskip
\bigskip


\maketitle

One of the crucial ingredients in the description of the nuclear
excitation in the continuum is the re-interaction between the emitted
nucleon, and the remaining nucleus.  The continuum random phase
approximation (CRPA) theory describes this effect, commonly called
final state interaction (FSI), as linear combination of particle-hole
and hole-particle excitations. Recently, we have developed a technique
to solve CRPA equations with finite-range interactions by considering,
without approximations, the excitation to the continuum \cite{don11}.
The application of our approach to medium-heavy nuclei produces
satisfactory descriptions of the experimental data.  The positions of
the peaks in the excitation of the electromagnetic giant dipole and
quadrupole resonances are well reproduced, even though the
widths of the resonances are too narrow and their heights too high.
There are strong indications that these problems are related to the
hypotheses underlying the RPA theory which is limited to consider
one-particle one-hole excitations only \cite{kam04}.

In this article, we study the ability of our CRPA calculations to
describe the electromagnetic responses of the $^4$He nucleus. The
application of the CRPA approach to the $^4$He nucleus is quite
unusual, since the number of particles composing the system is too
small to consider the mean-field hypotheses, on which the RPA theory
is based, to be reliable. On the other hand, for the $^4$He nucleus we
have the possibility of comparing our results with those of a fully
microscopic approach, based on the Lorentz Inverse Transform (LIT).
This approach uses nucleon-nucleon interactions constructed to
describe the two-nucleon systems and solves the Schr\"odinger equation
without approximations, contrary to the CRPA, which is an effective
theory where the many-body effects are considered by changing the
parameters of the interaction.

We have constructed the single particle basis by doing Hartree-Fock
calculations. We tested the validity of our description of the $^4$He
ground state, by using three effective interactions: two different
parameterizations of the Gogny interaction, the more traditional D1S
\cite{ber91} interaction and the more modern D1M force \cite{gor09},
which produces a reasonable neutron matter equation of state, and an
old finite-range effective interaction constructed to reproduce at
best the $^4$He binding energy, the B1 interaction of Brink and Boeker
\cite{bri67}.

While the D1S and D1M interactions have been widely used in the
literature to study the properties of medium-heavy nuclei, the
applications of the B1 interaction have been limited to the $^4$He and
to some test case for microscopic many-body calculations, since this
interaction has a soft core short-range repulsion in the scalar
channels. The properties of these two types of interactions are rather
different. For example, the D1S and D1M interactions reproduce the
empirical values of saturation density and binding energy per nucleon
of symmetric nuclear matter, while the symmetric nuclear matter
equation of state of the B1 interaction saturates at 0.21 fm$^{-3}$
with the energy per nucleon of -15.7 MeV. Also the values of the
symmetry energies, strictly related to the position of the peak of the
isovector dipole resonance, are rather different. For the D1M and D1S
interactions we obtain respectively the values of 29.45 MeV and 31.77
MeV, in general agreement with the commonly accepted empirical values
included in the ranges from 30 to 35 MeV. For the B1 interaction we
obtain the value of 58.55 MeV.

\begin{table}[htb]
\begin{center}
\begin{tabular}{lcccccccc}
\hline \hline 
&&&&\multicolumn{3}{c}{separation energies} &~~&  \\ \cline{5-7}
&~~~~& binding energy &~~~~& protons &~~& neutrons && rms \\\hline 
D1S &&  30.28 &&  19.39 &&  20.09 && 2.04\\ 
D1M &&  29.54 &&  18.25 &&  18.96 && 2.02\\ 
B1  &&  28.48 &&  26.00 &&  26.00 && 1.92\\ 
\hline 
exp &&  28.29 &&  19.81 &&  20.58 && 1.68\\ 
\hline \hline
\end{tabular}
\end{center} 
\vspace*{-.5cm}
\caption{\small Binding energies,  
  proton and neutron separation energies, in
  MeV, and rms charge radii, in fm,
  obtained for the three different interactions considered in
  this work. The experimental values are taken from Refs. 
  \cite{aud03,ang04a}.  }
\label{tab:energ}
\end{table}

The binding energies and the proton and neutron separation energies
obtained by using the three different interactions are given in Table
\ref{tab:energ}.  The experimental values have been taken from the
compilations of Refs. \cite{aud03,ang04a}. The performances of the HF
theory in the description of the $^4$He ground state properties are
quite unsatisfactory. The values of the binding energies generated by
the two Gogny interactions are too large with respect to the
experimental value. By construction, the B1 interaction makes a better
job in this case. The situation is reversed when the proton and
neutron separation energies are considered. In this case, the two
Gogny interactions provide a better description than the B1 force.

\begin{figure}[b]
\begin{center}
\includegraphics[scale=0.7]{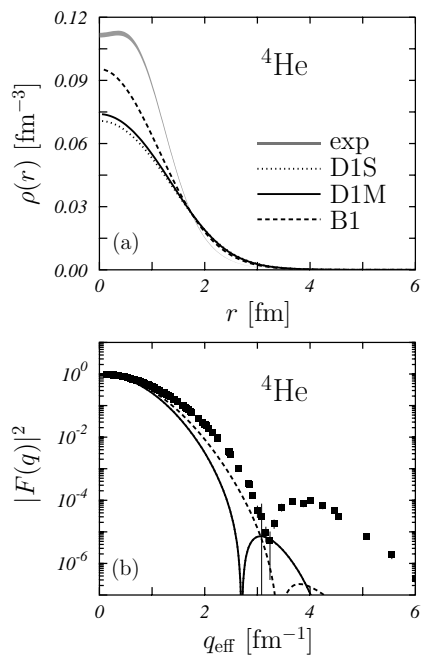} 
\end{center}
\vspace*{-.5cm}
\caption{\small (color on line). Panel(a): charge density
  distributions calculated with the D1S
  (dotted line) and D1M (solid line) parameterizations of the Gogny
  interaction and with the B1 interaction (dashed-dotted line) 
  compared to the empirical density taken from Ref. \cite{dej87}.
  Panel (b): form factors obtained with the charge distributions of
  the upper panel. The experimental data are rom Refs.
  \cite{fro67,eri68,mcc77,arn78}.}
\label{fig:gs}
\end{figure}

We compare in Fig. \ref{fig:gs}(a) our charge distributions with the
empirical one \cite{dej87}.  The discrepancies are remarkable
especially if compared with the good description of the charge
distributions of medium-heavy nuclei obtained by using the D1M and D1S
interactions \cite{don11}. In the present case, the charge
distributions are more extended than the experimental one as it is
also indicated by the values of the root mean squared (rms) charge
radii compared in Table \ref{tab:energ}.
We complete the information about the charge density distributions 
by comparing in Fig. \ref{fig:gs}(b) the elastic form factors obtained
with these charge densities with the experimental data of Refs. 
  \cite{fro67,eri68,mcc77,arn78}. The theoretical charge densities are
  larger than the empirical one and produce form factors which
  narrower than the experimental one. 

The results we have just presented confirm the difficulties of the
mean-field model in producing a good description of the $^4$He ground
state properties.  In any case, we are interested in investigating the
capacity of our self-consistent approach to describe the excitation of
the $^4$He nucleus in the continuum. 

\begin{figure}
\begin{center}
\includegraphics[scale=0.5]{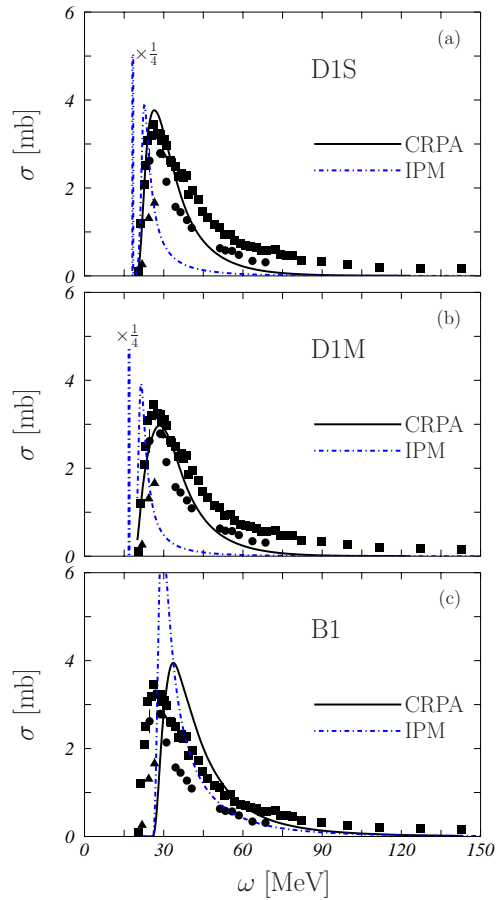} 
\end{center}
\vspace*{-.5cm}
\caption{\small
 Total photoabsorption cross sections obtained with the
 three interactions used in this work. The full lines show the results of
 the self-consistent CRPA calculations and the dashed-dotted  
 lines show the IPM results based on the HF calculations 
 done with the various interactions.
 The experimental data are from
 Refs. \cite{ark79} (squares), \cite{shi05} (triangles) and
 \cite{nil05} (circles). 
}
\label{fig:dipole}
\end{figure}

As a first test of the CRPA results we have calculated the total
photoabsorption cross section by using the method described in
  detail in Ref. \cite{don11} where it has been applied to some oxygen
  and calcium isotopes.  
We compare in Fig. \ref{fig:dipole} our
results with the available experimental data \cite{ark79,shi05,nil05}.
We have obtained the total cross section by summing the contribution
of the 1$^-$ and of the 2$^+$ excitations which contributes
to the total cross section only for about the 2\%. Panels (a), (b) and
(c) show the results obtained with the D1S, D1M and B1 interactions,
respectively. With the dashed-dotted lines we show the results of the
independent particle model (IPM) calculation, i.e. those obtained by
switching off the residual interaction in the CRPA calculation. The
solid lines show the CRPA results. 

The resonance is dominated by the transition of both protons and
neutrons from the $1s_{1/2}$ state to the $p$ waves, both $p_{3/2}$
and $p_{1/2}$. The IPM results indicate the resonances of these two
partial waves. In the D1S and D1M calculations the $1p_{3/2}$ state is
bound, while the $p_{1/2}$ state is in the continuum.  In the figure,
the bound responses for these two calculations, represented by the
narrow vertical lines in the panels (a) and (b) around 18 MeV, have
been multiplied by a 1/4 factor.  In the calculations with the B1 interaction only the $1s_{1/2}$
  states are bound, and already the IPM generates a peak of the
  resonance at energies higher than that of the experimental peak.
The inclusion of the residual interaction generates an additional
shift at even higher energies. 
In general, the residual interaction shifts the IPM
responses to higher energies, since the $1^-$ resonance is an
isovector excitation.

\begin{figure}
\begin{center}
\includegraphics[scale=0.6]{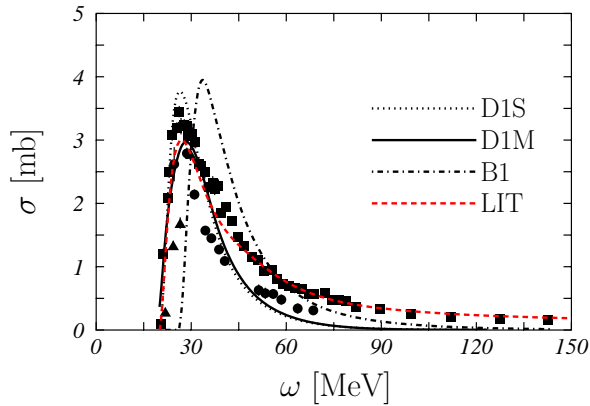} 
\end{center}
\vspace*{-.5cm}
\caption{\small 
 Comparison of the total photoabsorption cross sections
 obtained in the self-consistent CRPA calculations with the LIT
 results of Ref. \cite{gaz06}. The experimental data
 are from Refs. \cite{ark79,shi05,nil05}.}
\label{fig:dipole2}
\end{figure}

The direct comparison between our CRPA results, the experimental
  data and the results of of the microscopic calculation of
  Refs. \cite{gaz06} (dashed curve) based on the LIT technique is done
  in Fig. \ref{fig:dipole2}. This figure emphasizes the poor
  performances of the B1 interaction in reproducing the experimental
  data. This result confirms the indication already 
  given by the large value
  of the nuclear matter symmetry energy: the isospin part of the
  B1 interaction is too strong.  
More interesting is agreement
between the results obtained with the D1S and D1M interactions and
those of the microscopic calculation is remarkable.  
The peaks of
  the resonances generated by the three calculations are in the same
  position. In the peak, the D1M cross section has a better agreement
  with the LIT result with respect to the D1S one. This seems to
  indicate that the modifications of the parameters aimed to produce a
  reasonable neutron matter equation of state improved the description
  of the isospin channels of the interaction. In any case these
  differences are within the uncertainties of the input of the RPA
  calculations and not related to method itself. For this reason we
  perfom calculations with different interactions. We find more
  interesting the general differences between our RPA results and
  those of the LIT. 
The results of our calculations drop more
quickly in the high energy tail. Even though the experimental
situation is still quite controversial, the microscopic calculations
seem to give a better description of the data.It is worth to remark that, in our computational
  scheme \cite{don11}, ground state properties, single particle
  energies and wave functions and CRPA responses are stricly related 
  by the use of the same effective nucleon-nucleon interaction. 

\begin{figure}
\begin{center}
\includegraphics[scale=0.5]{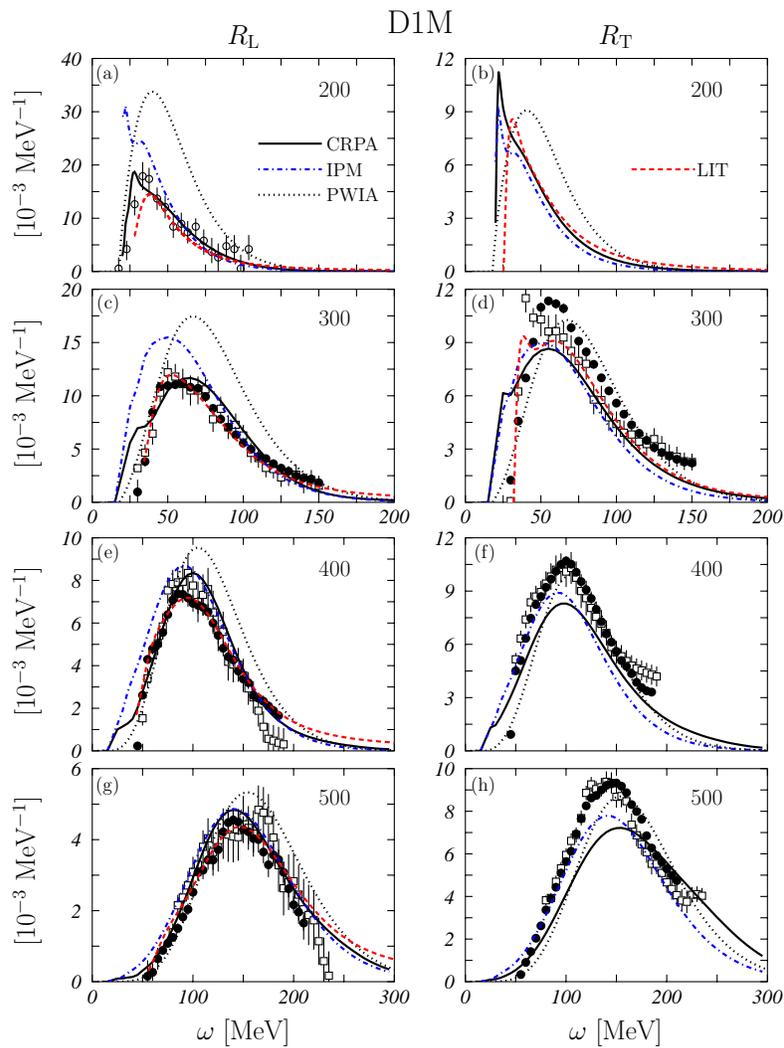}
\end{center}
\vspace*{-0.5cm}
\caption{\small 
Comparison of the longitudinal (left panels) and
transverse (right panels) quasi-elastic electron scattering responses
obtained with the D1M interaction within the CRPA (solid curves), IPM
(dashed-dotted curves) and PWIA (dotted curves) frameworks, with the
LIT microscopic results of Ref. \cite{bac07,bac09a,bac09b} (dashed curves). 
The labels in the panels indicate the values of the momentum
transfer in MeV/$c$.  The experimental data for momentum transfer of
200 MeV/$c$ (open circles) are taken from Ref. \cite{buk06} while
those shown in the other panels are taken from Ref. \cite{red90} (open
squares) and from Ref. \cite{zgh94} (solid circles).  }
\label{fig:qe}
\end{figure}

  We have tested the relevance of a good description of the charge
  density distribution in the evaluation of 1$^-$ excitation by using
  single particle wave functions generated by a Woods-Saxon potential
  whose parameters have been properly chosen.  The calculations have
  been done within a discretize RPA model, and compared with the
  analog self-consistent calculations done with HF wave functions. 
  The response strongly
  depends on the effective interaction, and no sensitive improvement
  with respect to the self-consistent approach has been found.  

We used our computational scheme to calculate the quasi-elastic
electron scattering responses.  
In these calculations we used the
one-body electromagnetic currents and the nucleonic electromagnetic
form factors described in Ref. \cite{ama93}.
We have done calculations with all the
three interactions mentioned before.  However, since the results are
rather similar, we present in Fig. \ref{fig:qe} only those results 
obtained with the D1M interaction.

The calculations of the quasi-elastic responses have been done by
summing the contribution of all the electric and magnetic multipole
excitations up to numerical convergence was reached. This has
been achieved by considering multipole excitations up to angular
momentum $J=6$ for the results at momentum transfer value $q=200$
MeV/c, and up to $J=8$ for $q=500$ MeV/c. In our calculations we have 
considered only one-body electromagnetic currents.

In Fig. \ref{fig:qe} the results of the CRPA calculations are
indicated by the full lines, while with the dashed-dotted curves we
show the IPM results. We have performed also calculations were the FSI
has been totally switched off, that is, we have calculated the
responses in IPM and we have substituted the mean-field wave functions
of the emitted nucleon with plane waves. We have indicated as plane
wave impulse approximation (PWIA) these results and we present them by
using dotted lines.  The dashed lines indicate the LIT results of
Refs.  \cite{bac07,bac09a,bac09b}.  The data for q=200 MeV/c are those
of Ref. \cite{buk06}, while for the other values of the momentum
transfer the black squares indicate the data of Ref. \cite{red90} and
the white circles those of Ref. \cite{zgh94}.

We observe first that the agreement between the CRPA results and the
experimental data is quite good for all the values of the momentum
transfer considered in the case of the longitudinal responses (left
panels). This confirms the results of the photoabsorption cross
section. In the case of the transverse response, our results 
slightly underestimate the data. In our calculations we did not
consider the meson-exchange-currents which enhance the transverse
response \cite{ang96,hal96}.

A second point is that CRPA effects become smaller with increasing
value of the momentum transfer in the longitudinal response. This can
be seen by comparing the CRPA results with those of the IPM. The full
and dashed-dotted lines are quite different for 200 and 300 MeV/c, but
they become closer at 400 MeV/c, and almost overlap at 500 MeV/c.

Transverse responses are more sensitive to the presence of CRPA
correlations. The differences with the IPM results grow slightly with
the momentum transfer. On the other hand, the comparison with the PWIA
results, the dotted lines, indicates that the mean field is taking
into account a large part of FSI for all the momentum transfer values
considered.  It is interesting to remark again the good agreement
between our CRPA results and those obtained with the fully microscopic
calculations done with the LIT technique, shown by dashed lines. 

While the HF theory gives a poor description of the $^4$He ground
state, the self-consistent CRPA theory describes well the excitation
of the continuum, for both photoabsorption and quasi-elastic inclusive
electron scattering data. We may say that our CRPA calculations are
able to describe well the FSI between the emitted nucleon and the
remaining nucleus. This good description of the FSI is obtained for a
wide range of values of the momentum transfer.

It is surprising that the performances of the CRPA are superior in
$^4$He than in medium-heavy nuclei, where the theory is supposed to be
tailored.  In medium-heavy nuclei a spreading width should be added to
have reasonable description of the excitation data in the
continuum. As it is shown in Ref. \cite{don11}, the difficulties of
the CRPA in describing the responses of medium-heavy nuclei are due to
the fact that excitations more complex than one-particle one-hole are
not considered. The effects of these excitations are almost absent in
$^4$He, and for this reason the CRPA works very well in this case.

\vskip 0.5 cm 
We thank S. Bacca, W. Leidemann and G. Orlandini for providing us with
the results of their calculations, and for the useful discussions.
This work has been partially supported by the Spanish Ministerio de
Ciencia e Innovacion under Contract Nos. FPA2009-14091-C02-02 and
ACI2009-1007 and by the Junta de Andaluc\'{\i}a (Grant No. FQM0220).


\clearpage
\newpage
\clearpage
\newpage
\end{document}